# Text Enhancement for Paragraph Processing in End-to-End Code-switching TTS

*Chunyu Qiang[1,2], Jianhua Tao[1,2,3], Ruibo Fu[1,2], Zhengqi Wen[1], Jiangyan Yi[1], Tao Wang[1,2], Shiming Wang[1,4]*

[1]National Laboratory of Pattern Recognition, Institute of Automation, Chinese Academy of Sciences, Beijing, China
[2]School of Artificial Intelligence, University of Chinese Academy of Sciences, Beijing, China
[3]CAS Center for Excellence in Brain Science and Intelligence Technology, Beijing, China
[4]School of Information Science and Technology, University of Science and Technology of China, Hefei, China
{chunyu.qiang,jhtao,ruibo.fu,zqwen,jyyi,tao.wang,shiming.wang}@nlpr.ia.ac.cn

## Abstract

Current end-to-end code-switching Text-to-Speech (TTS) can already generate high quality two languages speech in the same utterance with single speaker bilingual corpora. When the speakers of the bilingual corpora are different, the naturalness and consistency of the code-switching TTS will be poor. The cross-lingual embedding layers structure we proposed makes similar syllables in different languages relevant, thus improving the naturalness and consistency of generated speech. In the end-to-end code-switching TTS, there exists problem of prosody instability when synthesizing paragraph text. The text enhancement method we proposed makes the input contain prosodic information and sentence-level context information, thus improving the prosody stability of paragraph text. Experimental results demonstrate the effectiveness of the proposed methods in the naturalness, consistency, and prosody stability. In addition to Mandarin and English, we also apply these methods to Shanghaiese and Cantonese corpora, proving that the methods we proposed can be extended to other languages to build end-to-end code-switching TTS system.

## 1. Introduction

With the development of deep learning, the end-to-end speech synthesis method has been proposed such as Char2Wav [1] and Tacotron [2]. Code-switching refers to the process of switching the linguistic code from one to another, used to synthesis the text including other language words or phrases. A code-switching TTS system should be able to generate high-quality speech, and it is also considered the same speaker when generating speech of another language, also known as consistency.When a code-switching TTS system is constructed from the corpora of different speakers, the naturalness and consistency will be poor. And when synthesizing paragraph text, the prosody instability will appear. However, there are few research works on end-to-end TTS method to solve these problems.

The statistical parameter speech synthesis (SPSS) methods use a corpora recorded by a bilingual speaker to build a bilingual TTS system [3]. A Chinese-English bilingual TTS system also constructed from such corpora, which uses different front-ends to process text in different languages, and synthesize the text with a single voice [4]. This method records a multilingual speech corpora by speakers who are proficient in multiple languages. It needs to develop a phoneme set that contains phonemes in multiple languages. SPSS based on the hidden Markov models (HMMs) is also used to construct the code-switching TTS system [5, 6, 7]. The mapping is learned through bilingual corpora recorded by a bilingual person. But it is hard to obtain multilingual speech corpora by speakers who are proficient in multiple languages. To solve this problem, the voice conversion method was used to create a multilingual corpora from a set of multilingual speech corpora, and the code-switching TTS can synthesize the target speaker speech in four languages [8]. However, this method is limited by the effect of voice conversion. To handle these problems, mixed-coding method is applied to the code-switching TTS system. A TTS framework with mixed-coding writes text in scripts of other languages [9, 10]. Then, the TTS system is trained on the monolingual corpora using the mapping within phonemes of multiple languages. A bilingual TTS system was constructed on two monolingual speech corpora using the mixed-coding method [11]. Recently, end-to-end speech synthesis method is used to construct the code-switching TTS system [12]. A code-switched end-to-end TTS system was proposed with consistent voice using only monolingual corpora [13]. An end-to-end code-switching TTS system with cross-lingual word embedding was proposed, to improve the voice rendering [14].

The existing mixed-coding method cannot solve the problem of consistency when the speakers of the bilingual corpora are different, because the similar syllables of different languages are independent of each other. In this paper, we propose a cross-lingual embedding layers structure to solve the problem of naturalness and consistency. There are some prosody instability when synthesizing paragraph text. The method of adding extra contextual information to improve the performance of prosody stability has also been applied to end-to-end speech synthesis [13]. We propose a text enhancement method that makes the input contain prosodic information and sentence-level context information to solve the problem of paragraph text prosody instability. In addition to Mandarin and English, we also apply these methods to Shanghaiese and Cantonese corpora, proving that the methods we proposed can be extended to other languages to build end-to-end code-switching TTS system.

In this paper, our contributions can be summarized as follows. (1) We propose a Cross-lingual Embedding layers structure. The method can improve the naturalness and consistency of generated code-switching speech with multi-speaker bilingual corpora. (2) We propose a text enhancement method makes the input contain prosodic information and sentence-level context information. The method can improve the prosody stability of paragraph text. (3) Our proposed method is language independent, which can be extended to other languages to build end-to-end code-switching TTS system.

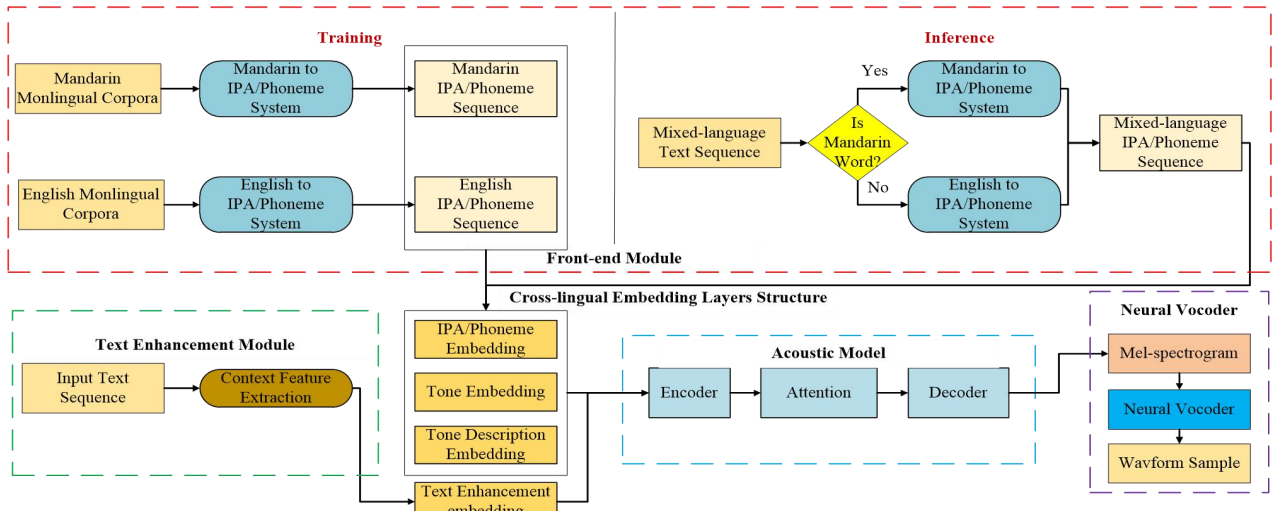

Figure 1: *End-to-end code-switching TTS framework. These dotted frames represent the front-end module, text enhancement module, acoustic model, and neural vocoder. These embedding layers constitute the cross-lingual embedding layers structure.*

## 2. Methods

As Figure 1 shows, the end-to-end code-switching TTS system consists of four components. The red dotted frame contains the front-end module for converting Mandarin or English sentence into IPA or phoneme sequence. It can generate both tone embedding and tone description embedding. The green dotted frame contains the text enhancement module for generating context information embedding layers, such as prosody boundary, word-level context information, sentence-level context information. These embedding layers constitute the cross-lingual embedding layers structure. The blue dotted frame contains the acoustic model (based on Tacotron). It is an encoder-decoder architecture with the attention mechanism which predicts Bark-scale cepstral coefficients and pitch parameters from IPA sequences. The purple dotted frame contains the neural vocoder (based on LPCNet) which predicts waveform samples from Bark-scale cepstral coefficients and pitch parameters.

### 2.1. Cross-lingual Embedding Layers Structure

In this paper, we built a cross-lingual embedding layers structure on the Mandarin and English corpora. The method we proposed can be extended to other languages in the features to build end-to-end code-switching TTS system.

#### 2.1.1. Cross-lingual Embedding of Phoneme and Tone

Mandarin characters are ideograms and cannot represent sound. Considering there are thousands of Mandarin characters, it is impossible to use Mandarin characters as the input of end-to-end TTS system. To reduce the size of the Mandarin characters set, the phoneme is chosen to replace the character input, which is extremely close to the Latin character set in number. In the English end-to-end TTS system, the main method takes letter or phoneme as input. At present, whether in Mandarin or English, the use of phonemes as input can be synthesized into higher quality speech. However, it is still out of the question to use phonemes as input when building a code-switching cross-lingual embedding layer structure. Since the Mandarin is a Sino-Tibetan language which is monosyllabic and tonal, while English is atonal and has stress rules. When the phoneme set with tone in Mandarin and the phoneme set with stress in English are mixed together, the phoneme set will be too large. Consequently, we use three embedding layers to describe pronunciation when using the phoneme labeling method.

As Figure 2 shows, the three embedding layers contain phoneme (without tone) embedding layer, tone embedding layer and tone description embedding layer. In this solution, Mandarin phonemes without tones and English phonemes with stress are used separately as phoneme (without tone) embedding layer. This greatly reduces the sparsity of the phoneme distribution. Tone is essential to the description of pronunciation, so we added the tone embedding layer. To describe the tone more fully, we added a tone description embedding layer. Since each tone has fluctuations, we describe this fluctuation through a series of sequences, for instance, /3 4 5/ is used to describe the second tone of Mandarin syllable.

The cross-lingual embedding layers structure of using phoneme we proposed can construct a high-quality end-to-end code-switching TTS while using a bilingual speech corpora recorded by a bilingual speaker. However, if the data is not from the same speaker, it will have trouble in the consistency and naturalness of speech.

#### 2.1.2. Cross-lingual Embedding of IPA and Tone

Because of uncorrelation between phonemes in Mandarin and English, and syllable labeling with the same pronunciation are different, the model is difficult to represent the consistency. We propose a cross-lingual embedding layer structure based on IPA.

As Figure 2 shows, the two embedding layers contain phoneme (without tone) embedding layer and tone embedding layer. In this solution, Mandarin IPA without tones and English IPA with stress are used separately as IPA (without tone) embedding layer. In order to associate Mandarin and English syllable labeling, IPA is split into single character for one-hot encoding. On the tone embedding layer, the construction rules are similar to the structure of using phoneme. Since almost all languages can be described by IPA, this method can be extended to other languages.

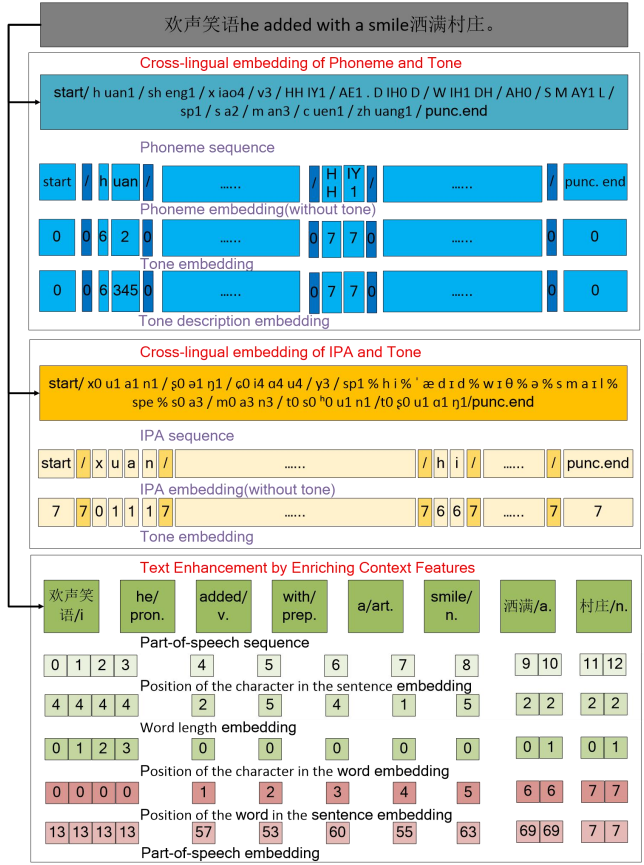

Figure 2: *Cross-lingual Embedding layers structure . Example "欢声笑语洒满村庄" means "the village is full of laughter"*

### 2.2. Text Enhancement by Enriching Context Features

*2.2.1. Text Enhancement Method*

Although the methods we proposed can generate high-quality speech, there is still a problem of poor prosody performance when generating paragraph text. Unlike English, there is no clear break between Mandarin words, so this problem is more serious in Mandarin. The way of using IPA or phonemes as input cannot contain context features, but when actual human speaking, the context has a great influence on the performance of speech. Therefore, we enhance the stability and naturalness of paragraph text code-switching TTS by text enhancement. Although speech-text data is difficult to obtain, the text data is easier to obtain. So we use a large amount of text data to build the text enhancement module.

As Figure 1 shows, a text enhancement module is added before the encoder to generate multiple text information vectors. When training or inferencing, the prosodic boundary generated by the text enhancement model is added to the IPA or phoneme sequence, then concatenate the IPA or phoneme embedding layer with text enhancement embedding layers. Figure 2 describes text enhancement embedding in a sentence. Based on word segmentation and part-of-speech tagging, we generated multiple context text features embedding layers.

*2.2.2. Features Choices*

We found that the stability and naturalness of speech have been improved by text enhancement method. We want to research which combination of context features is the most effective. As shown in Table 1, we select the following features as input:

- The prosody boundary information includes prosodic word boundary, prosodic phrase boundary and international phrase boundary. These features apply to ITE-B system, ITE-BW system, ITE-BS system and ITE-BWS system.

- The word-level context information includes the length of the current word, the part-of-speech of the current word and the position of the current character in the word. These features apply to ITE-BW system and ITE-BWS system.

- The sentence-level context information includes the position of the current character in sentence and the position of the current word in sentence. These features apply to ITE-BS system and ITE-BWS system.

## 3. Experiments and Results Analysis

In the experiments, We compare the differences between the existing method(PE system) and the proposed method, and evaluate the effectiveness of the several combinations of code-switching cross-lingual embedding layers structure approaches in the consistency and naturalness of synthesized speech. In addition, we evaluate the effectiveness of several text enhancement methods to improve the performance of prosodic phrases when synthesizing paragraph text.

### 3.1. Corpora and Features

We evaluate the proposed methods on Mandarin and English dataset recorded different speakers. We leverage several online implementation packages: the py2ipa package [17], the g2pE package [18] and the eng_to_ipa package [19] to build our front-end module.

- Mandarin: a dataset with 10,000 sentences recorded by a woman whose native language is Mandarin: 9000 sentences as training set, 500 sentences as validation set, and the rest 500 sentences are reserved as test set.

- English: a dataset with 7,000 sentences recorded by a woman whose native language is English: 6,300 sentences as training set, 350 sentences as validation set, and the rest 350 sentences are reserved as test set.

In this paper, all audios are down-sampled to 24kHz. The extracted acoustic features are 32-dimensional: including 30 dimensional Bark-scale cepstral coefficients [20] and 2 dimensional pitch parameters.

Table 1: *Combination of cross-lingual embedding layers and context features.*

| Systems | cross-lingual embedding Layers | | | | Context Features | | |
|---|---|---|---|---|---|---|---|
| | Phoneme embedding | IPA embedding | Tone embedding | Tone description embedding | Prosody boundary | Word-level context | Sentence-level context |
| PE | ● | ○ | ○ | ○ | ○ | ○ | ○ |
| IE | ○ | ● | ○ | ○ | ○ | ○ | ○ |
| PTE | ● | ○ | ● | ● | ○ | ○ | ○ |
| ITE | ○ | ● | ● | ○ | ○ | ○ | ○ |
| ITE-B | ○ | ● | ● | ○ | ● | ○ | ○ |
| ITE-BW | ○ | ● | ● | ○ | ● | ● | ○ |
| ITE-BS | ○ | ● | ● | ○ | ● | ○ | ● |
| ITE-BWS | ○ | ● | ● | ○ | ● | ● | ● |

*Note: In the PE system and the IE system, the IPA embedding or phoneme embedding contain tones, and IPA is not split into single character.*

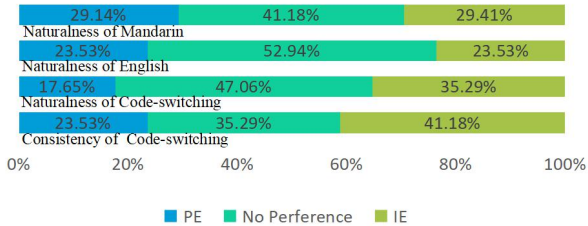

Figure 3: *AB preference results of PE and IE.*

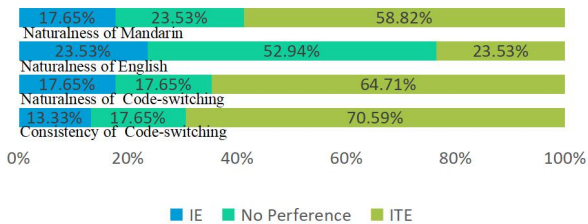

Figure 4: *AB preference results of IE and ITE.*

### 3.2. Experimental Setting

In the training of acoustic models, the frame length is 40ms and the frame shift is 10ms. The adam optimization algorithm [21] and learning rate attenuation mechanism are applied. The adam coefficient is 0.999 and the initial learning rate is 0.0005. The Stop-token is also applied to the model to dynamically stop decoding. The model is trained to about 300,000 steps, and the batch size is 32. In the training of LPCNet, the AMSGrad optimization algorithm [22] is applied, the model is trained to about 120 steps, and the batch size is 64. All experiments are completed on Tensorflow [23] . As shown in Table 1, the architecture describes the cross-lingual embedding layers structure and the context features in each systems. The first four systems are used to compare the effects of the cross-lingual embedding layers structure. And the last four systems are used to compare which context feature is most effective.

### 3.3. Results of Methods

#### 3.3.1. Evaluation of cross-lingual embedding Layers Structure

We perform AB preference tests in terms of naturalness and consistency to assess the performances of different methods. The consistency AB preference tests are conducted on code-switching sentences. Each set of 20 sentences is randomly selected from test set. A group of 14 subjects were asked to choose which one was better in terms of the naturalness and consistency of synthesis speech.

The percentage preference is shown in Figure 3 and Figure 4. We can see that the ITE system can achieve better naturalness of Mandarin and code-switching, and can achieve better consistency of code-switching. However, the IE system is not improved relative to PE system. Using only IPA sequences as input is similar to phoneme sequences as input, the same syllables of Mandarin and English are independent of each other. Thus there is no significant difference between the results of IE system and PE system. The IPA labeling and embedding layers construction method we proposed lead to the same syllable related, while different languages tone embedding have subtle discrepancy. Therefore, the ITE system performs best and its naturalness of English is the same as IE system.

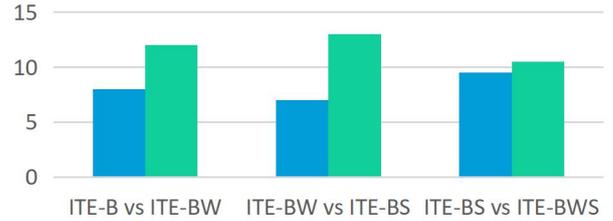

Figure 5: *System preference analysis.*

#### 3.3.2. Evaluation of Enhancement Methods

We conducted three sets of comparative (ITE-B vs. ITE-BW; ITE-BW vs. ITE-BS; ITE-BS vs. ITE-BWS). Each set of 20 paragraph text is randomly selected from test set. A group of 14 subjects were asked to choose which one was better in terms of the prosody stability. We use paired-samples t-tests to examine the differences between each of the two systems.

The result of t-test is shown in Figure 5. ITE-BW system performs better than ITE-B system [$p<0.001$], ITE-BS system performs better than ITE-BW system [$p<0.001$], ITE-BWS system and IBW system behave similarly [$p = 0.0043$]. Experimental results prove that the text enhancement methods by the prosody boundary and sentence-level context information is most effective.

## 4. Conclusions

In this paper, we propose an end-to-end code-switching TTS framework, including the cross-lingual embedding layers structure and the text enhancement method. The ITE system makes the same syllable in different language relevant. The cross-lingual embedding layers structure can effectively improve the naturalness and the consistency of code-switching. The ITE-BS system makes the input contain prosodic boundary and sentence-level positional context information, which in line with human speech. The text enhancement method can solve the problem of long code-switching sentences prosody instability. The experimental results demonstrate the effectiveness of the proposed methods. In addition to Mandarin and English, we also apply these methods to Shanghaiese and Cantonese corpora, proving that the methods we proposed can be extended to other languages to build end-to-end code-switching TTS system.

In future research, we will try to improve our proposed methods in other languages. Besides, we will also try to use a single language corpora to build a code-switching TTS system that can synthesize speech in other languages.

## 5. Acknowledgements

This work is supported by the National Key Research & Development Plan of China (No.2017YFB1002801), the National Natural Science Foundation of China (NSFC) (No.61831022, No.61901473, No.61771472, No.61773379, No.61771472, No.61773379 )